\documentstyle[aps]{revtex}
\begin{document}
\draft
\title{Transition to Fulde-Ferrel-Larkin-Ovchinnikov phases near the 
tricritical point : an analytical study }
\author{R. Combescot and C. Mora}
\address{Laboratoire de Physique Statistique,
 Ecole Normale Sup\'erieure*,
24 rue Lhomond, 75231 Paris Cedex 05, France}
\date{Received \today}
\maketitle

\begin{abstract}
We explore analytically the nature of the transition to the Fulde-Ferrel-
Larkin-Ovchinnikov superfluid phases in the vicinity of the tricritical 
point, where these phases begin to appear. We make use of an 
expansion of the free energy up to an overall sixth order, both in order 
parameter amplitude and in wavevector. We first explore the 
minimization of this free energy within a subspace, made of arbitrary 
superpositions of plane waves with wavevectors of different 
orientations but same modulus. We show that the standard second order 
FFLO phase transition is unstable and that a first order transition occurs 
at higher temperature. Within this subspace we prove that it is favorable 
to have a real order parameter and that, among these states, those with 
the smallest number of plane waves are prefered. This leads to an order 
parameter with a $\cos({\bf  q}_{0}. {\bf  r})$ dependence, in 
agreement with preceding work. Finally we show that the order 
parameter at the transition is only very slightly modified by higher 
harmonics contributions when the constraint of working within the 
above subspace is released.
\end{abstract}

\section{INTRODUCTION}
Although they have been proposed a long time ago, Fulde-Ferrel-
Larkin-Ovchinnikov (FFLO) phases \cite{ff,larkov}  are still the subject 
of a continuing interest. Indeed the existence of these phases is a fairly 
remarkable phenomenon since they correspond to a spontaneous 
symmetry breaking of the standard BCS superfluid phase in the 
presence of an effective field, inducing a difference in chemical potential 
between the two populations involved in the formation of Cooper pairs. 
This symmetry breaking leads to an inhomogeneous superfluid with a 
space dependent order parameter, while the applied field is perfectly 
homogeneous. This situation is analogous to the appearance of vorticity 
in type II superconductors, but in this latter case the effect is due to the 
coupling of the field to particle currents while in FFLO phases only the 
coupling to the spin of the pairing fermions is involved. In standard 
superconductors the coupling to the orbital degrees of freedom is much 
stronger than the coupling to the spins. Hence the upper critical field is 
due to the orbital coupling and the FFLO phases can not be observed, 
since they should appear at much higher field. However in heavy 
fermions superconductors the strength of these two couplings is 
comparable, which could make possible the observation of FFLO 
phases. Nevertheless their sensitivity to impurities could be a major 
problem. Another possible direction to eliminate the orbital coupling is 
to consider lower dimensional superconductors, in a geometry where 
the currents would have to flow in an actually prohibited direction. 
Organic compounds or cuprate superconductors are interesting systems 
in this respect. And indeed very recently the FFLO state has been 
claimed to be observed in a quasi-two-dimensional organic compound 
\cite{singleton} . On the other hand earlier possible observations in 
heavy fermion compounds \cite{gloos} have not been undisputed. We 
note in particular that the analysis of experimental results relies very 
often heavily on the theoretical results, but we will see that the situation 
is not completely satisfactory in this respect.

\vspace{4mm} 
Another class of physical systems where FFLO phases could be 
observed is coming up quite recently. These are the ultracold fermionic 
gases. As it is well known remarkably low temperatures have been 
obtained on bosonic gases, leading in particular to the observation of 
Bose-Einstein condensation in alkali ultracold gases. More recently 
fermionic gases have been cooled down in the degenerate regime 
\cite{marcojin,trusc,salomon} and reaching a BCS superfluid transition 
in these systems seems a reasonable possibility \cite{stoofal,rc} . 
However in the systems considered for observing this transition there 
is, in contrast to electronic spin relaxation in superconductors, no fast 
relaxation mechanism to equalize the populations of the two fermion 
species involved in the formation of Cooper pairs. Hence one should 
have no limitation to the effective field in these systems since the 
number of atoms in the populations can be in principle obtained at will. 
So the difference in atomic populations looks as a very promising 
control parameter. On the other hand if this parameter is not fully 
controlled this might very well be a major difficulty in reaching the BCS 
transition in these systems \cite{bkk,rcffan} .

\vspace{4mm} 
In contrast with this raising experimental interest there are still 
theoretical problems with the precise nature of the possible phases. 
Specifically the basic FFLO instability corrresponds to have pairs 
formed with a total nonzero momentum ${\bf  q}_{0}$ instead of 
forming pairs (${\bf  k}, -{\bf  k}$) with zero total momentum as in the 
standard BCS phase. This gives rise to a spatial dependence $\exp(i{\bf  
q}_{0}. {\bf  r})$ for the order parameter, which leaves a degeneracy 
with respect to the orientation of $ {\bf  q}_{0}$. This has been 
investigated by \cite{larkov}  Larkin and Ovchinnikov (LO) who 
looked how it is lifted right below the critical field. In this case, when 
considering the spatial dependence of the order parameter $\Delta ( {\bf  
r})$, one can restrict the investigation to the subspace generated by 
linear combination of the plane waves $\exp(i{\bf  q}_{0}. {\bf  r})$ 
with all possible directions ${\bf  q}_{0}$. At T = 0, LO looked for 
periodic structures and found that the energetically favored result is a 
second order transition to a one-dimensional 'planar' texture $\Delta ( 
{\bf  r}) \sim \cos({\bf  q}_{0}. {\bf  r})$. However they left open in 
their paper the possibility of a first order transition. Actually when 
considering the three-dimensional 'cubic' texture $\Delta ( {\bf  r}) \sim 
\cos({q}_{0}x) + \cos({q}_{0}y) + \cos({q}_{0}z) $ they found that 
it is energetically unfavorable compared to the normal state. 
Nevertheless they obtained from the gap equation that a nonzero 
solution for this order parameter exists \emph{above} the FFLO 
transition line. In terms of the expansion of the free energy in powers of 
the order parameter, schematically $ \Omega = \alpha _{2} \Delta^{2} 
+ \alpha _{4} \Delta^{4}  $, this situation corresponds to a positive 
coefficient $\alpha _{2}$ for the second order term and a negative 
coefficient $\alpha _{4}$ for the fourth order one, just the opposite of 
the standard Landau-Ginzburg expansion below the transition. The LO 
evaluation of the related free energy corresponds actually to the 
maximum $ \alpha _{2}^{2}/ 4 | \alpha _{4} | $ of this free energy. 
Beyond this maximum the free energy decreases, and it would go to $ - 
\infty $ if one would consider only the second and fourth order terms. 
Naturally one has to include the effect of all higher order terms in order 
to find the value of the free energy for large values of the order 
parameter. However, at the FFLO transition line and slightly above it, 
the free energy becomes negative for values of the order parameter 
(specifically for $\Delta^{2} =  \alpha _{2} / | \alpha _{4} | $) where 
only the second and fourth order terms have to be kept in the 
expansion. This shows definitely that the transition toward the FFLO 
phase occurs \emph{above} the standard second order FFLO transition 
line and is actually first order, because one can display in this range a 
solution which has a lower free energy than the normal state. On the 
other hand in order to obtain consistently the order parameter which 
gives the lowest free energy one has naturally to take into account 
higher order terms. In particular it is by no means obvious that the cubic 
phase is the stable one. Moreover when higher order terms are 
considered there are no reason anymore to restrict the search to the 
subspace generated by the plane waves $\exp(i{\bf  q}_{0}. {\bf  r})$.

\vspace{4mm} 
In order to explore more fully this difficult problem with easier 
conditions, it is better to be able to proceed to some kind of expansion. 
This can be done if, instead of working at  $ T = 0 $, one explores the 
vicinity of the tricritical point (TCP), where the FFLO transition line 
starts. It is located at  $T_{\rm{tcp}} / T_{c0} = 0.561 $ where 
$T_{c0}$ is the critical temperature for  $ \bar{ \mu } = 0 $ , with $ 2 
\bar{ \mu } = \mu _{\uparrow} - \mu_{\downarrow} $ being half the 
chemical potential difference between the two fermionic populations 
forming pairs. The corresponding effective field is  $ \bar{ \mu 
}_{\rm{tcp}}/ T_{c0} = 1.073 $. At this point, in the free energy 
expansion, both the coefficient of the second ordre term $ \alpha _{2}$ 
and the coefficient of the fourth ordre term $ \alpha _{4}$ vanish. 
Naturally  $ \alpha _{2}$ is  zero just because the TCP is on the 
standard second order phase transition line. On the other hand 
following this transition line one has $ \alpha _{4}>0 $ for $ T > 
T_{\rm{tcp}}$ and $ \alpha _{4}< 0 $ for $ T < T_{\rm{tcp}}$. The 
change of sign of $ \alpha _{4} $ at $ T_{\rm{tcp}}$ is the origin of 
the FFLO instability. Clearly, by continuity, the equilibrium order 
parameter will be small in the vicinity of the TCP since we know that it 
is zero just above the TCP on the second order line and it is small right 
below it in the superfluid phase. Therefore a power expansion of the 
free energy will be enough to find it. On the other hand, since the 
second and fourth order terms are zero at the TCP, we have clearly to 
expand at least up to sixth order, but this will prove to be enough 
because the corresponding coefficient is positive and not small. 
Similarly the optimum wavevector ${\bf  q}_{0}$ corresponding to the 
FFLO phase will be small in the vicinity of the TCP since it is zero just 
above it on the second order line. This allows to proceed to a gradient 
expansion of the free energy. Since only even powers of the 
wavevector can enter and we look for a minimum as a function of this 
wavevector, we have to expand at least up to fourth order in gradient, 
but this will prove again to be enough. 

\vspace{4mm} 
This line of thought has actually already been followed by Houzet et al. 
\cite{buz1,buz2}, who have performed this expansion for the free 
energy and explored the result numerically. They have found that the 
energetically favored phase is the one-dimensional planar order 
parameter found by LO at $ T = 0 $, but that the transition is actually 
slightly first order, instead of second order as found by LO at $ T = 0 
$. Our purpose in the present paper is rather to proceed to an analytical 
study of this problem. Indeed there are infinitely many possible order 
parameters in competition. And our aim, in considering the vicinity of 
the TCP, is to find the important ingredients which are responsible for 
the selection of the actual stable state and obtain a better physical 
understanding, having in particular in mind the generalization to more 
complicated situations. Hence our paper is complementary to their 
work. In particular we obtain a first order transition to the one-
dimensional planar order parameter, but we will be able to analyze the 
reasons which favor this phase. The transition to this planar order 
parameter has been actually explored numerically down to $ T = 0 $ by 
Matsuo et al. \cite{matsuo} . They have used quasiclassical equations 
and found that the transition keeps first order down to low temperature, 
but eventually goes to second order in agreement with LO. On the other 
hand since we know that  at $ T = 0 $ the cubic phase is more stable 
than the planar one, the question of the stablest phase at low 
temperature is still unsolved. Finally this first order transition to the 
planar phase in the three-dimensional case is in contrast with the results 
of Burkhardt and Rainer \cite{br} who found it to be second order in a 
two-dimensional space.
 
\vspace{4mm} 
In the following section, for completeness and to set up our notations, 
we rederive the expression \cite{buz2} of the free energy. After 
considering in section III some simple situations, we explore in details 
the minimization of this free energy. This is done in section IV by 
restricting our search to the LO subspace for the order parameter, which 
is made of arbitrary superpositions of plane waves. In section V we 
show that our results are only slightly modified when we release this 
restriction. Throughout the paper we restrict ourselves to the simplest 
BCS scheme, namely we will consider the free energy corresponding to 
a weak coupling isotropic Fermi system, ignoring in particular any 
Fermi liquid effect. Moreover we concentrate on the three-dimensional 
case which leads to a first order transition, and make only occasionally 
comparison with the two-dimensional situation where the transition is 
second order.

\section{THE  FREE  ENERGY}

There are various ways to obtain the explicit expression for the free 
energy we need in the vicinity of the TCP  \cite{agd,eil}. In practice it 
is convenient to use the fact \cite{eil}  that, by varying the free energy 
with respect to $\Delta ( {\bf  r})$ one finds the gap equation, which is 
also easily obtained from GorkovÕs equations, as done for example by 
LO \cite{larkov} . The integral form of these equations is 
\cite{gork,agd} , with standard notations:
\begin{eqnarray}
G _{\uparrow}( {\bf  r},{\bf  r'}) = G^{0} _{\uparrow}( {\bf  r} - 
{\bf  r'}) -
\int d{\bf  r}_{1} G^{0} _{\uparrow}( {\bf  r} - {\bf  r}_{1}) \Delta 
({\bf  r}_{1})
F ^{+}( {\bf  r}_{1},{\bf  r'})
\label{eq1}
\end{eqnarray}
\begin{eqnarray}
F ^{+}( {\bf  r},{\bf  r'}) =
\int d{\bf  r}_{1} \bar{G^{0} _{\downarrow}}( {\bf  r} - {\bf  
r}_{1}) \Delta ^{*}({\bf  r}_{1})
G _{\uparrow} ( {\bf  r}_{1},{\bf  r'})
\label{eq2}
\end{eqnarray}
with, for the Fourier transforms of the free fermions thermal 
propagators, $ G^{0} _{\uparrow}( {\bf k}) = (i \omega _{n} - \xi 
_{{\bf  k}} + \bar{ \mu }) ^{-1}$ and $ \bar{G^{0} _{\downarrow}} 
( {\bf k}) = (-i \omega _{n} - \xi _{k} - \bar{ \mu }) ^{-1}$ where  
$\xi _{ {\bf  k}}$ is the kinetic energy measured from the Fermi 
surface for $ \bar{ \mu } = 0 $ and $ \omega _{n} = \pi T (2n+1) $ are 
Matsubara frequencies. The order parameter is given by the self-
consistency relation :
\begin{eqnarray}
\Delta ^{*}({\bf  r}) = V T \sum_{n} F ^{+}( {\bf  r},{\bf  r})
\label{eq3}
\end{eqnarray}
We expand Eq. (1-2) up to fifth order in  $\Delta ({\bf  r})$. We 
introduce the Fourier transform $\Delta _{{\bf  q}} = \int   d{\bf  r} 
\Delta ({\bf  r}) \exp(-i{\bf  q}.{\bf  r})$ of the order parameter. As 
explained in the introduction  we proceed also to an expansion in the 
wavevector ${\bf  q}$ of the order parameter since we know that its 
relevant values will be small in the vicinity of the TCP. More precisely 
we will see that, in order to obtain a coherent expansion, it is enough to 
go only up to fifth order terms in overall power of $ \Delta $ and ${\bf  
q}$. This means that, in the gap equation, we have to expand the first 
order term in $ \Delta $ only to fourth order in ${\bf  q}$. Similarly the 
third order term in $ \Delta $ has to be expanded only to second order in 
${\bf  q}$ and the fifth order term in $ \Delta $ can be calculated to 
zeroth order in ${\bf  q}$. For example in order to find the third order 
term, we have to expand up to second order in wavevectors :
\begin{eqnarray}
\sum_{k}  \bar{G^{0}}( {\bf k}) G^{0} ( {\bf k}+{\bf q}_{1})  
\bar{G^{0}}( {\bf k} + {\bf q}-{\bf q}_{3}) G^{0} ( {\bf k}+{\bf 
q}) \Delta ^{*} _{{\bf  q}_{1}} \Delta _{{\bf  q}_{2}} \Delta ^{*} 
_{{\bf  q}_{3}}
\label{eq4}
\end{eqnarray}
where we have used ${\bf  q}_{1} + {\bf  q}_{3} = {\bf  q} + {\bf  
q}_{2}$ and we have omitted the unnecessary spin index. In the 
expansion appear the following numerical coefficients:
\begin{eqnarray}
& a _{0}(\bar{\mu },T) =  \frac{1}{N _{0}V} - 2 \pi T  \: \Re [ 
\sum_{n=0} \frac{1}{\bar{\omega  }_{n}}] & \nonumber
\end{eqnarray}
\begin{eqnarray}
& a _{2}(\bar{\mu }/T) = - \bar{\mu } ^{2} 2 \pi T \: \Re [ \sum_{ 
n=0} \frac{1}{\bar{\omega  }_{n} ^{3}} ] &
\end{eqnarray}
\begin{eqnarray}
& a _{4}(\bar{\mu }/T) = - \bar{\mu } ^{4} 2 \pi T \: \Re [ \sum_{ 
n=0} \frac{1}{\bar{\omega  }_{n} ^{5}} ] \nonumber
\label{eq5}
\end{eqnarray}
where $\bar{\omega  }_{n} = \omega _{n} - i \bar{\mu }$, and the 
summation for $ a _{0}$ has to be cut-off in the standard BCS way. 
The simple second order transition line to a standard BCS superfluid 
with space independent order parameter is given by  $ a _{0}(\bar{\mu 
},T) = 0 $. Below the TCP it corresponds to a spinodal transition line, 
at which the normal state becomes absolutely unstable against a 
transition toward a space independent order parameter. The domain $ a 
_{0}(\bar{\mu },T) > 0 $ corresponds to the region of the $ (\bar{\mu 
},T) $ phase diagram above this line, and it is the domain where we 
will look for other transitions. In practice we can see $ a _{0}(\bar{\mu 
},T) $ as a measure of the distance from the spinodal line in the $ 
(\bar{\mu },T) $ plane. Explicitely if we define $ T _{sp}(\bar{\mu 
}/T)$ the spinodal temperature as a function of the ratio $\bar{\mu }/T 
$, we have $ a _{0}(\bar{\mu },T) = \ln [T/ T _{sp}(\bar{\mu }/T)]$.  
We will not need to explicit further this distance. As indicated in the 
introduction we have by definition $ a _{2}(\bar{\mu }/T) = 0 $ at the 
TCP and it is small in the vicinity of this point. For $ (\bar{\mu }/T) > 
(\bar{\mu }/T) _{\rm{tcp} } = 1.91 $, we have $ a _{2} > 0 $ and $ a 
_{2} < 0 $ for $ (\bar{\mu }/T) < (\bar{\mu }/T) _{\rm{tcp}}$. 
Finally $ a _{4}(\bar{\mu }/T) = 0.114 $ at the TCP (while it is 
negative near $ \bar{\mu } = 0 $ and goes to $ -0.25$ when $ T 
\rightarrow 0 $). With these notations the gap equation in the vicinity of 
the TCP reads :
\begin{eqnarray}
\Delta _{{\bf  q}} [ a _{0} - \frac{1}{3} a _{2} Q ^{2} + \frac{1}{5} 
a _{4} Q ^{4} ] 
 -  \sum_{{\bf  q}_{i}}  \Delta _{{\bf  q}_{1}} \Delta ^{*} _{{\bf  
q}_{2}} \Delta _{{\bf  q}_{3}} [\frac{1}{2} a _{2} -  \frac{1}{6} a 
_{4} (Q ^{2} + 2 Q ^{2} _{1} + 2 Q ^{2} _{3} - {\bf  Q}.( {\bf  
Q}_{1} + {\bf  Q}_{3} ) + 3 {\bf  Q}_{1}.{\bf  Q}_{3})]  \nonumber   
\\
 + \frac{3}{8} a _{4} \sum_{{\bf  q}_{i}}  \Delta _{{\bf  q}_{1}} 
\Delta ^{*} _{{\bf  q}_{2}} \Delta _{{\bf  q}_{3}}  \Delta ^{*} 
_{{\bf  q}_{4}} \Delta _{{\bf  q}_{5}} = 0
\label{eq6}
\end{eqnarray}
where we have used the dimensionless wavevector $ {\bf  Q} = {\bf  
q} v _{F} / 2 \bar{\mu }$ and expressed $  \Delta _{{\bf  q}} $ in 
units of $ \bar{\mu }$. Also the momentum conservation is assumed in 
the summations, that is ${\bf  q}_{1} + {\bf  q}_{3} = {\bf  q} + {\bf  
q}_{2}$ in the third order term and ${\bf  q}_{1} + {\bf  q}_{3} + 
{\bf  q}_{5} = {\bf  q} + {\bf  q}_{2} + {\bf  q}_{4}$ in the fifth 
order one. The above expression can be checked against the case of the 
simple Fulde-Ferrell state $\Delta ( {\bf  r}) = \exp(i{\bf  q}_{0}. {\bf  
r})$ where a single wavevector enters.

\vspace{4mm} 
Now the above gap equation is obtained by minimizing the following 
free energy difference $ \Omega $ between the superfluid and the 
normal state :
\begin{eqnarray}
\Omega  = \sum_{{\bf  q}}  | \Delta _{{\bf  q}} |^{2} [ a _{0} - 
\frac{1}{3} a _{2} Q ^{2} + \frac{1}{5} a _{4} Q ^{4} ] 
 - \frac{1}{2} \sum_{{\bf  q}_{i}}  \Delta _{{\bf  q}_{1}} \Delta 
^{*} _{{\bf  q}_{2}} \Delta _{{\bf  q}_{3}}  \Delta ^{*} _{{\bf  
q}_{4}} [\frac{1}{2} a _{2} -  \frac{1}{6} a _{4} ( Q ^{2}_{1} + Q 
^{2} _{2} +  Q ^{2} _{3} + Q ^{2} _{4} + {\bf  Q}_{1}.{\bf  
Q}_{3} + {\bf  Q}_{2}.{\bf  Q}_{4})]  \nonumber   \\
+ \frac{1}{8} a _{4} \sum_{{\bf  q}_{i}}  \Delta _{{\bf  q}_{1}} 
\Delta ^{*} _{{\bf  q}_{2}} \Delta _{{\bf  q}_{3}}  \Delta ^{*} 
_{{\bf  q}_{4}} \Delta _{{\bf  q}_{5}}  \Delta ^{*} _{{\bf  q}_{6}}
\label{eq7}
\end{eqnarray}
where we have the momentum conservation ${\bf  q}_{1} + {\bf  
q}_{3} = {\bf  q}_{2} + {\bf  q}_{4}$ in the fourth order term while 
${\bf  q}_{1} + {\bf  q}_{3} + {\bf  q}_{5} = {\bf  q}_{2} + {\bf  
q}_{4} + {\bf  q}_{6}$ holds in the sixth order one. We have used 
symmetry and momentum conservation to present the fourth order term 
in a symmetrical way. This expression Eq.(7) is just the free energy we 
were looking for. We have considered here the 3-D case. For a two-
dimensional system the angular averages found in the calculation are 
different. The result is simply obtained from the above one by 
multiplying the $ Q ^{2}$ terms by $3/2$ and the $ Q ^{4}$ terms by 
$15/8$.

\section{SIMPLE CASES}

Let us first consider some simple situations. If we consider an 
homogeneous order parameter, that is $ {\bf  q} = 0 $, we have merely 
$ \Omega  = a _{0} \Delta^{2}- a _{2} \Delta^{4}/4 + a _{4} 
\Delta^{6}/8 $ . If we want to have this free energy negative for $ a 
_{0} > 0 $ we need to have $ a _{2} > 0 $, that is to be at temperature 
below the TCP. In this case $ \Omega > 0 $ when $ a _{0} > a 
_{2}^{2} / 8 a _{4} $ and we reach a first order transition for $ a _{0} 
= a _{2}^{2} / 8 a _{4} $, with a non zero order parameter  $ 
\Delta^{2} = a _{2} /  a _{4} $. This is the standard first order Pauli 
limiting transition. We consider next the possibility of a second order 
transition. In this case only the second order term in Eq.(7) is relevant. 
The location of this transition is given by $ a _{0} = \frac{1}{3} a 
_{2} Q ^{2} - \frac{1}{5} a _{4} Q ^{4} $. Below the TCP, where $ 
a _{2} > 0 $, we can find $ a _{0} > 0 $ for non zero wavevector $ 
{\bf  Q}$, that is we will find an FFLO phase. Precisely the optimal 
wavevector is $ Q_{0} ^{2} = \frac{5}{6} a _{2} /  a _{4}$ and the 
corresponding maximal  $ a _{0} $ is  $ a _{0} =  \frac{5}{36} a 
_{2}^{2} / a _{4} $. We see that this value is larger than the one we 
just found for the standard Pauli limiting transition. Thus as expected 
the FFLO transition happens first and overtakes the first order 
transition. Finally it is natural and interesting to try to generalize the two 
above situations and consider the possibility of a first order transition 
for an order parameter with a single wavector component $ \Delta 
_{{\bf  q}}$. With the shorthand $ \Delta _{{\bf  q}} \equiv \Delta $ 
the free energy writes :
\begin{eqnarray}
\Omega  =  [ a _{0} - \frac{1}{3} a _{2} Q ^{2} + \frac{1}{5} a _{4} 
Q ^{4} ] \Delta ^{2}
- \frac{1}{4} [ a _{2} -  2 a _{4} Q ^{2} ] \Delta ^{4} + \frac{1}{8} a 
_{4} \Delta ^{6}
\label{eq8}
\end{eqnarray}
Minimizing first with respect to $ Q ^{2} $ we obtain for the extremum 
the condition $ Q ^{2} = \frac{5}{6} a _{2} /  a _{4} - \frac{5}{4} 
\Delta ^{2} $, which implies that must have $ \Delta ^{2} \leq 
\frac{2}{3} a _{2} /  a _{4}$ otherwise we are back to the 
homogeneous situation and the Pauli limiting transition. Inserting this 
value for $ Q ^{2}$ in Eq.(8) we find $ \Omega  =  [ a _{0} -  
\frac{5}{36} a _{2}^{2} / a _{4}] \Delta ^{2}+ \frac{1}{6} a _{2} 
\Delta ^{4} - \frac{3}{16} a _{4} \Delta ^{6}$. We are naturally 
interested in finding a transition higher than the standard FFLO. This 
means we are looking for $ a _{0} >  \frac{5}{36} a _{2}^{2} / a 
_{4} $, so the first term in $ \Omega $ is positive. But one sees that the 
sum of the last two terms is also positive for $ \Delta ^{2} \leq 
\frac{2}{3} a _{2} /  a _{4}$. Therefore we have not been able to 
improve the standard FFLO solution. However we have clearly not 
done our best in this direction.

\vspace{4mm} 
Before trying to improve in this way, it is convenient to simplify our 
expression for the free energy by taking reduced units for the order 
parameter and the wavector, which come out naturally from our above 
discussion. We set $ \Delta = (a _{2} /  a _{4}) ^{1/2} \bar{\Delta}$, $ 
{\bf  Q} =  (a _{2} /  a _{4}) ^{1/2} \bar{{\bf  q}}$ , $ a _{0} = A 
_{0} a _{2}^{2} / a _{4}$ and $ \Omega = (a _{2}^{3} / a _{4}^{2}) 
F $. This leads to rewrite Eq.(7) for the free energy as :
\begin{eqnarray}
F  = \sum_{{\bf  q}}  | \bar{\Delta} _{{\bf  q}} |^{2} [ A _{0} - 
\frac{1}{3} \bar{q} ^{2} + \frac{1}{5} \bar{q}^{4} ] 
-  \sum_{{\bf  q}_{i}} \bar{\Delta} _{{\bf  q}_{1}} \bar{\Delta} ^{*} 
_{{\bf  q}_{2}} \bar{\Delta} _{{\bf  q}_{3}} \bar{\Delta} ^{*} 
_{{\bf  q}_{4}} [\frac{1}{4} -  \frac{1}{12} (\bar{q} ^{2}_{1} + 
\bar{q} ^{2} _{2} +  \bar{q} ^{2} _{3} + \bar{q} ^{2} _{4} +  
\bar{{\bf  q}}_{1}. \bar{{\bf  q}}_{3} +  \bar{{\bf  q}}_{2}. 
\bar{{\bf  q}}_{4})]  \nonumber  \\
+ \frac{1}{8} \sum_{{\bf  q}_{i}}  \bar{\Delta} _{{\bf  q}_{1}} 
\bar{\Delta} ^{*} _{{\bf  q}_{2}}  \bar{\Delta} _{{\bf  q}_{3}}   
\bar{\Delta} ^{*} _{{\bf  q}_{4}}  \bar{\Delta} _{{\bf  q}_{5}}  
\bar{\Delta} ^{*} _{{\bf  q}_{6}}
\label{eq9}
\end{eqnarray}
It is clear from this rescaling transformation and the resulting 
expression that, in the vicinity of the TCP (where $ a _{2}$ is small), it 
is unnecessary to go beyond our sixth order expansion in $ \Delta $ and 
$ {\bf  q}$. It is also of interest to rewrite this free energy as a 
functional of $ \Delta ({\bf  r})$ by Fourier transform. This gives, after 
by parts integrations :
\begin{eqnarray}
F  = \int  d{\bf  r} [A _{0}  | \bar{\Delta} |^{2} -  \frac{1}{3}  | {\bf 
\nabla} \bar{\Delta} |^{2} + \frac{1}{5}  | {\bf \nabla}^{2} 
\bar{\Delta} |^{2}] 
-  \int  d{\bf  r} [\frac{1}{4}  | \bar{\Delta} |^{4} - \frac{1}{24} [ 2 
({\bf \nabla} | \bar{\Delta} |^{2}) ^{2} + 3  ({\bf \nabla} 
\bar{\Delta}^{2})  ({\bf \nabla} \bar{\Delta}^{*2})]]  +   \frac{1}{8}  
\int  d{\bf  r}  | \bar{\Delta} |^{6}
\label{eq10}
\end{eqnarray}

\section{THE   LO   SUBSPACE}

As emphasized by LO all the states corresponding to the same wavector 
$\bar{q}_{0}$ but with different orientation for $\bar{\bf{q}}_{0}$ 
are degenerate right on the FFLO transition line. This degeneracy is 
lifted, at least partially, when one goes into the superfluid phase 
because of the coupling between the various plane waves produced by 
the nonlinear terms in the free energy. When one investigates the states 
selected in this process, one has to consider the subspace:
\begin{eqnarray}
\bar{\Delta} ({\bf  r}) = \sum \Delta _{{\bf  q}_{0}} \exp(i{\bar{\bf  
q}}_{0}. {\bf  r})
\label{eq11}
\end{eqnarray}
of all the order parameters generated by these plane waves. We call this 
the LO subspace. In this section we will restrict to this subspace our 
search for the state appearing at the transition : we will look  for the 
minimum of the free energy within this LO subspace. Actually LO 
looked for a lattice as a solution and restricted themselves to this kind of 
order parameter. However there is physically no basic reason to enforce 
this type of restriction. One could look for incommensurate structures 
or quasicrystal-like solutions. Even if these are not the lowest energy 
solution, they might be of interest as local solutions corresponding 
physically to defects. Therefore we have not set a periodicity condition 
on the solutions we have considered. Nevertheless let us indicate at 
once that the energetically favored solutions we have found within the 
LO subspace are actually periodic. Although considering only the LO 
subspace is an important restriction, it does not make the problem easy 
at all, although we solve it completely below.

\vspace{4mm} 
Let us first show that, within this subspace, the standard FFLO 
transition line is not stable, and the transition is actually first order. 
With our reduced units the LO subspace corresponds to 
$\bar{\bf{q}}_{0}^{2} = \frac{5}{6}$. This minimizes the coefficient 
of the second order term in Eq.(9). The FFLO transition line is then 
given by $ A _{0} = \frac{5}{36}$, which makes the second order 
term zero within the LO subspace. Let us then look at the fourth order 
term in Eq.(9). This amounts to calculate the functional derivative of the 
free energy on the FFLO transition line. In a standard second order 
phase transition it should always be positive, forcing the order 
parameter to be zero on the transition line. However in the present case 
it is not obvious that this is systematically so because of the interplay of 
the wavevectors in this term. Specifically we introduce a parameter $ 
\beta $ which describes this effect for any fixed order parameter $\Delta 
({\bf  r})$. It is defined by :
\begin{eqnarray} 
2 \beta  \bar{q}_{0}^{2} \sum_{{\bf  q}_{i}} \bar{\Delta} _{{\bf  
q}_{1}} \bar{\Delta} ^{*} _{{\bf  q}_{2}} \bar{\Delta} _{{\bf  
q}_{3}} \bar{\Delta} ^{*} _{{\bf  q}_{4}} =   \bar{q}_{0}^{2} 
\sum_{{\bf  q}_{i}} ( \hat{{\bf  q}}_{1}. \hat{{\bf  q}}_{3} +  
\hat{{\bf  q}}_{2}. \hat{{\bf  q}}_{4})  \bar{\Delta} _{{\bf  q}_{1}} 
\bar{\Delta} ^{*} _{{\bf  q}_{2}} \bar{\Delta} _{{\bf  q}_{3}} 
\bar{\Delta} ^{*} _{{\bf  q}_{4}} = -  \int   \bar{\Delta}^{2}  ({\bf 
\nabla} \bar{\Delta}^{*})^{2} + c.c.
\label{eq12}
\end{eqnarray}
For the simple case of the FF solution we have merely $ \beta  = 1$. 
However this is the highest possible value and we can think of 
decreasing it, or even making it negative, by a proper choice of the 
order parameter in the LO space, although naturally $ \beta  \ge  -1$. 
Then the fourth order term in Eq.(9) is given by $[ \frac{1}{6} ( \beta 
+ 2 ) \bar{q}_{0}^{2} - \frac{1}{4} ]  \int  | \bar{\Delta} |^{4} $ and 
has thus the same sign as $ 5 \beta + 1 $ when we take $  
\bar{q}_{0}^{2} =  \frac{5}{6}$.  Hence for any order parameter 
with $ \beta \le - \frac{1}{5} $ the fourth order term is negative. We 
will find actually many such states. Now when we are on the FFLO 
transition line (the second order term is zero) and the fourth order term 
is negative, we decrease the free energy and make it negative just by 
taking the order parameter to be small and nonzero, which shows that 
the standard FFLO transition line is unstable. Naturally the larger the 
order parameter, the lower the free energy, but we have to stay in the 
range where the second and fourth order terms are the only ones 
important in our expansion, which means that the sixth order term is 
negligible. Then since we have a negative free energy, we can raise it 
back to zero by increasing $ A _{0}$ beyond its value $ \frac{5}{36}$ 
on the FFLO line, which means we go beyond this line in the $ 
(\bar{\mu },T) $ phase diagram. In this way the second order term 
becomes positive. Eventually we will be limited by the growth of the 
sixth order term. The transition we have found corresponds to positive 
second and sixth order terms and a negative fourth order one. The free 
energy becomes negative for a nonzero value of the order parameter. 
We have thus found a first order transition beyond the standard FFLO 
transition line. This discussion about the order of the transition is the 
exact analogue of the one we made in the introduction for the $T = 0$ 
situation.

\vspace{4mm} 
Naturally it is of interest to minimize $ \beta $ since it is rather natural to 
expect that the states corresponding to the minimum will lead to the 
stronger instability toward the first order transition. We show now that 
$ \beta \geq -\frac{1}{3} $, the equality $ \beta = -\frac{1}{3} $ being 
obtained for \textit{any} real order parameter. We make use of :
\begin{eqnarray}
\int  [ \bar{\Delta}^{2}  ({\bf \nabla} \bar{\Delta}^{*})^{2} -  | 
\bar{\Delta} |^{2} | {\bf \nabla} \bar{\Delta} |^{2}] + c.c. = \int  [ 
\bar{\Delta} {\bf \nabla} \bar{\Delta}^{*} - c.c.] ^{2} \leq 0
\label{eq13}
\end{eqnarray}
where the equality occurs only for a real order parameter (within an 
irrelevant overall constant phase factor), and : 
\begin{eqnarray}
\int  [ \bar{\Delta}^{2}  ({\bf \nabla} \bar{\Delta}^{*})^{2} +  2 | 
\bar{\Delta} |^{2} | {\bf \nabla} \bar{\Delta} |^{2}] + c.c. = - \int   | 
\bar{\Delta} |^{2} (\bar{\Delta} {\bf \nabla}^{2} \bar{\Delta}^{*} + 
c.c.) = 2 \bar{q}_{0}^{2} \int | \bar{\Delta} |^{4}
\label{eq14}
\end{eqnarray}
where the last step makes specific use of the form Eq.(11) for the order 
parameter. We have then :
\begin{eqnarray}
- 2 \beta \bar{q}_{0}^{2} \int | \bar{\Delta} |^{4} = \int   
\bar{\Delta}^{2}  ({\bf \nabla} \bar{\Delta}^{*})^{2} + c.c. \leq \int  [ 
\bar{\Delta}^{2}  ({\bf \nabla} \bar{\Delta}^{*})^{2} - \frac{2}{3} [ 
\bar{\Delta}^{2}  ({\bf \nabla} \bar{\Delta}^{*})^{2} -  | \bar{\Delta} 
|^{2} | {\bf \nabla} \bar{\Delta} |^{2}] ] + c.c. \\ \nonumber
= \frac{1}{3} \int  [ \bar{\Delta}^{2}  ({\bf \nabla} 
\bar{\Delta}^{*})^{2} +  2 | \bar{\Delta} |^{2} | {\bf \nabla} 
\bar{\Delta} |^{2}] + c.c. = \frac{2}{3} \bar{q}_{0}^{2} \int | 
\bar{\Delta} |^{4}
\label{eq15}
\end{eqnarray}
hence $ \beta \geq -\frac{1}{3} $.

\vspace{4mm} 
Together with $ \beta $ it is also intuitively convenient to consider $ 
\gamma $ defined by : 
\begin{eqnarray} 
\gamma   \int d{\bf  r}  | \bar{\Delta} |^{4}  =   \bar{q}_{0}^{-2}  \int 
d{\bf  r} [{\bf \nabla} | \bar{\Delta} |^{2}] ^{2} = \sum_{{\bf  
q}_{i}} ( \hat{{\bf  q}}_{1} -  \hat{{\bf  q}}_{2} )^{2}  \bar{\Delta} 
_{{\bf  q}_{1}} \bar{\Delta} ^{*} _{{\bf  q}_{2}} \bar{\Delta} 
_{{\bf  q}_{3}} \bar{\Delta} ^{*} _{{\bf  q}_{4}}
\label{eq16}
\end{eqnarray}
which is easily seen to satisfy $ \gamma  = 1 - \beta $ from Eq.(14). 
We are thus interested in maximizing $ \gamma $. From the first 
expression in Eq.(16) it is intuitively clear that $ \gamma $ will be large 
when the (properly normalized) order parameter has strong spatial 
variations. In particular it will be better for $  | \bar{\Delta}({\bf  r}) 
|^{2}$ to have many nodes. This is more easily achieved if 
$\bar{\Delta}({\bf  r})$ has no imaginary part since in this case one has 
only to require that the real  part is zero. This makes intuitively 
reasonable that $ \gamma $ is maximized by a real order parameter. One 
can also come to this conclusion from the Fourier expansion in Eq.(16) 
(or also from Eq.(12)) : it is of interest to have as often as possible 
opposite wavevectors so that $  ( \hat{{\bf  q}}_{1} -  \hat{{\bf  
q}}_{2} )^{2} $ takes as much as possible its maximum value, namely 
4 . One is thus naturally led to an order parameter which is a 
combination of $\cos({\bf  q}_{0}. {\bf  r} + \varphi _{{\bf  q}_{0}}  
)$ with real coefficients, the directions of the ${\bf  q}_{0}$ 's being 
free. This corresponds merely to require that $  \bar{\Delta}({\bf  r}) $ 
is real in Eq.(11) by taking $ \Delta _{-{\bf  q}_{0}} = \Delta 
^{*}_{{\bf  q}_{0}}$. On the other hand, since any real order 
parameter gives the maximum $ \gamma $, it is not necessary that these 
cosines have equal weight.

\vspace{4mm} 
We can now come back to our free energy Eq.(9) and find the best 
solution within the LO subspace, since we have found that the 
minimum $ \beta $ is $ - 1/3 $, as soon as the order parameter is real, 
which implies $ \Delta _{-{\bf  q}_{0}} = \Delta ^{*}_{{\bf  
q}_{0}}$ in Eq.(11). We introduce a measure $ \bar{\Delta}$ of the 
amplitude of the order parameter by setting :
\begin{eqnarray}
\int  d{\bf  r} | \bar{\Delta} |^{2} = \sum_{{\bf  q}_{i}} \bar{\Delta} 
_{{\bf  q}_{1}} \bar{\Delta} ^{*} _{{\bf  q}_{2}} = N _{2} 
\bar{\Delta}^{2}
\label{eq17}
\end{eqnarray}
where by definition $ N _{2} \equiv N $ is the number of plane waves 
coming in Eq.(11). If all the planed waves have same amplitude, $ 
\bar{\Delta}$ is just the common value of these amplitudes. Then we 
define $ N _{4}$ and $ N _{6}$ by :
\begin{eqnarray}
\int | \bar{\Delta} |^{4} = \sum_{{\bf  q}_{i}} \bar{\Delta} _{{\bf  
q}_{1}} \bar{\Delta} ^{*} _{{\bf  q}_{2}} \bar{\Delta} _{{\bf  
q}_{3}} \bar{\Delta} ^{*} _{{\bf  q}_{4}} = N _{4} 
\bar{\Delta}^{4}
\label{eq18}
\end{eqnarray}
and :
\begin{eqnarray}
\int | \bar{\Delta} |^{6} = \sum_{{\bf  q}_{i}}  \bar{\Delta} _{{\bf  
q}_{1}} \bar{\Delta} ^{*} _{{\bf  q}_{2}}  \bar{\Delta} _{{\bf  
q}_{3}}   \bar{\Delta} ^{*} _{{\bf  q}_{4}}  \bar{\Delta} _{{\bf  
q}_{5}}  \bar{\Delta} ^{*} _{{\bf  q}_{6}} = N _{6} 
\bar{\Delta}^{6}
\label{eq19}
\end{eqnarray}
For the simple plane wave considered in Eq.(8), we had $ N _{2} = N 
_{4} = N _{6} = 1  $. For the real order parameter we are interested in, 
the set of wavevectors $\{ {\bf  q}_{i} \}$ is made of N/2 pairs. If all 
the planed waves have same amplitude, one finds \cite{note} for 
example by simple counting that $ N _{4} = 3 N (N-1) $ and $ N _{6}   
=  5 N ( 3 N ^{2} - 9 N  + 8 ) $. Actually, once it is recognized that, 
from the counting procedure, $ N _{4}$ and $ N _{6}$ are 
polynomials in $N$ of order 2 and 3 respectively, the coefficient may 
easily be found by considering the cases $N = 2 , 4 , 6 $. With these 
notations the free energy Eq.(9) reduces to : 
\begin{eqnarray}
F  = N _{2} \bar{\Delta}^{2} [ A _{0} - \frac{1}{3} \bar{q} ^{2} + 
\frac{1}{5} \bar{q}^{4} ]
- N _{4} \bar{\Delta}^{4} [\frac{1}{4} - \frac{ \alpha }{2} \bar{q} 
^{2} ]
+ \frac{1}{8} N _{6} \bar{\Delta}^{6}
\label{eq20}
\end{eqnarray}
where we have set $ \alpha = \frac{ \beta + 2}{3}$ (for the simple 
plane wave considered in Eq.(8), we had $ \alpha = 1$).

\vspace{4mm} 
We proceed now as we have done for Eq.(8). Minimizing F with 
respect to $  \bar{q} ^{2} $ we find for the extremum the condition 
$\bar{q} ^{2} = \frac{5}{6} - \frac{5}{4} \alpha (N _{4} / N _{2})  
\bar{\Delta} ^{2} $, which implies that, in our considerations, we have 
for $  \bar{\Delta} ^{2} $ an upper bound $ \bar{\Delta} _{max} ^{2} 
= (2/3 \alpha ) N _{2} / N _{4}$. This leads, for the value of the free 
energy F at this extremum, to :
\begin{eqnarray}
\frac{F}{ \bar{\Delta}^{2}} = N _{2}  ( A _{0} -  \frac{5}{36} ) + N 
_{4} \bar{\Delta}^{2} (\frac{5 \alpha }{12}- \frac{1}{4})+  
\bar{\Delta}^{4} (\frac{N _{6}}{8} - \frac{5 \alpha ^{2}}{16} \frac{ 
N _{4}^{2}}{ N _{2}})
\label{eq21}
\end{eqnarray}
Now the standard FFLO solution corresponds to $ A _{0} =  
\frac{5}{36} $. Since we are interested in a better solution we want $ A 
_{0} >  \frac{5}{36} $ which makes the first term of  $ F /  
\bar{\Delta}^{2} $ positive. On the other hand, for $ \bar{\Delta} =  
\bar{\Delta} _{max}$, the sum of the last two terms in the right hand 
side (r.h.s.) of Eq.(21) can be written as $ \bar{\Delta} _{max} ^{2} 
(N _{4} / 24 \alpha ) [ 5 ( \alpha - \frac{3}{5}) ^{2} + 2 N _{2} N 
_{6}/ N _{4} ^{2} - \frac{9}{5} ]$. This is always positive since we 
have $ N _{2} N _{6}/ N _{4} ^{2} \geq 1 $ ( this results directly 
from $ \int  d{\bf  r}  | \bar{\Delta} |^{2} . \int  d{\bf  r}  | \bar{\Delta} 
|^{6} \geq [ \int  d{\bf  r}  | \bar{\Delta} |^{4} ] ^{2} $). If we assume 
that the second term in the r.h.s. of Eq.(21) is positive, this implies that 
the free energy is always positive. Therefore if we want to find non 
positive values for F it is necessary to have a negative coefficient for the 
second term in the r.h.s. of Eq.(21), which means $ \alpha < 
\frac{3}{5}$. This is possible since the minimum $ \beta  = -
\frac{1}{3}$ we have found above corresponds to a minimum $ \alpha 
= \frac{5}{9} $.

\vspace{4mm} 
Then the quadratic form Eq.(21) can be zero if we meet the condition $ 
4 ( A _{0} -  \frac{5}{36} ) ( 2 \frac{ N _{2} N _{6}}{ N _{4} 
^{2}} - 5 \alpha ^{2} )  \leq  (\frac{5 \alpha }{3} - 1 ) ^{2} $. This 
leads to the following result for the transition line to the FFLO phase :
\begin{eqnarray}
A _{0} = \frac{5}{36} + \frac{1}{8}\: \frac{(1 - \frac{5 \alpha }{3} ) 
^{2}}{\frac{ N _{2} N _{6}}{ N _{4} ^{2}} - \frac{5}{2} \alpha 
^{2}}  
\label{eq22}
\end{eqnarray}
In the $ ( \bar{ \mu } - T )$ plane this line is higher than the standard 
second order FFLO transition line, which is given by  $ A _{0} =  
\frac{5}{36} $. On the other hand the value of $ \bar{\Delta}_{m}$ 
which gives $ F = 0 $ at the threshold given by Eq.(21) is : 
\begin{eqnarray}
\bar{\Delta}^{2}_{m} = \frac{ N _{2}}{ N _{4}}\: \frac{1 - \frac{5 
\alpha }{3}}{\frac{ N _{2} N _{6}}{ N _{4} ^{2}} - \frac{5}{2} 
\alpha ^{2}} 
\label{eq23}
\end{eqnarray}
which is clearly nonzero and our transition is quite explicitely a first 
order transition (note that the condition $ \bar{\Delta}_{m} < 
\bar{\Delta} _{max}$ is necessarily satisfied since F is zero for $ 
\bar{\Delta}_{m}$ and positive for $ \bar{\Delta} _{max}$ ). 

\vspace{4mm} 
Coming back to the location of the transition line Eq.(23) we consider 
now how to optimize it. First we see from the numerator that it is 
advantageous to lower $ \alpha $ as much as possible. Indeed our 
minimum $ \alpha = \frac{5}{9} $ is quite close to the limiting value $ 
\alpha = \frac{3}{5}$ so the variation of the denominator with $ \alpha 
$ is irrelevant. Hence we are lead to make the fourth order term in the 
free energy Eq.(9) as negative as possible by minimizing $ \beta $, as 
we anticipated at the beginning of section IV. Once we have taken $ 
\alpha = \frac{5}{9} $, we see that it is of interest to take $ N _{2} N 
_{6}/ N _{4} ^{2} $ as small as possible (we have seen that it is 
bounded from below by $1$). From their definitions Eq.(18) and 
Eq.(19) we can evaluate $ N _{4}$ and $ N _{6}$ for a general order 
parameter Eq.(11) $ \bar{\Delta} ({\bf  r}) = \sum \Delta _{{\bf  
q}_{i}} \exp(i{\bf  q}_{i}. {\bf  r}) = 2 \sum |\Delta _{{\bf  q}_{i}}| 
\cos({\bf  q}_{i}. {\bf  r} + \varphi _{{\bf  q}_{i}}  )$. One finds : 
\begin{eqnarray}
N _{4} \bar{\Delta}^{4} = 3 (\sum _{i} |\Delta _{{\bf  q}_{i}}| 
^{2})^{2} -  3 \sum _{i} |\Delta _{{\bf  q}_{i}}| ^{4}
\label{eq24}
\end{eqnarray}
and : 
\begin{eqnarray}
N _{6} \bar{\Delta}^{6} = 15 (\sum _{i} |\Delta _{{\bf  q}_{i}}| 
^{2})^{3} - 45 (\sum _{i} |\Delta _{{\bf  q}_{i}}| ^{2})^{2}\sum 
_{i} |\Delta _{{\bf  q}_{i}}| ^{4}+ 40 \sum _{i} |\Delta _{{\bf  
q}_{i}}| ^{6}
\label{eq25}
\end{eqnarray}
Actually one can recognize that, for symmetry reasons, the results 
involve only $ \sum _{i} |\Delta _{{\bf  q}_{i}}| ^{n}$ with $n = 2, 4, 
6$, because odd powers of cosines average to zero. Hence the results 
assume necessarily the general form given by Eq.(24) and Eq.(25). 
Then the coefficients are easily obtained from the specific case where all 
the amplitudes $  |\Delta _{{\bf  q}_{i}}|$ are equal.

\vspace{4mm} 
Since the wavevectors are paired it is now more convenient to sum over 
pairs from now on. Defining $ a _{i} = |\Delta _{{\bf  q}_{i}}| 
^{2}/\sum _{i}|\Delta _{{\bf  q}_{i}}| ^{2}$ (implying $ \sum _{i} a 
_{i} = 1 $) we have :
\begin{eqnarray}
\frac{9}{10}  N _{2} N _{6}/ N _{4} ^{2} = \frac{6  - 9 S _{2} + 4 
S _{3}}{(2  - S _{2}) ^{2}} 
\label{eq26}
\end{eqnarray}
where $ S _{2} =  \sum _{i} a _{i} ^{2}$ and  $ S _{3} =  \sum _{i} 
a _{i} ^{3}$. When we have a single pair $ S _{2} = S _{3} = 1 $ and 
the r.h.s. of Eq.(26) is equal to 1. We show now that it is otherwise 
larger than 1. Since $ (2  - S _{2}) ^{2} < 4 - 3 S _{2}$ because $ S 
_{2} < 1 $ when we have more than a single pair, it is enough to prove 
that $ 3 (1  - S _{2}) \geq 2  (1  - S _{3})$. This is in turn verified 
because we can write for the left-hand side $ 1  - S _{2} = (\sum _{i} a 
_{i}) ^{2} - \sum _{i} a _{i} ^{2} =  2 \sum _{i < j} a _{i} a _{j}$. 
In a similar way we have in the right-hand side $ 1  - S _{3} = (\sum 
_{i} a _{i}) ^{3} - \sum _{i} a _{i} ^{3} =  3 \sum _{i < j} a _{i} a 
_{j} (a _{i}+ a _{j}) + 6  \sum _{i < j < k} a _{i} a _{j} a _{k} = 3 
\sum _{i < j} a _{i} a _{j} - 9 \sum _{i < j < k} a _{i} a _{j} a _{k}$. 
So our statement is correct since it is just equivalent to $ \sum _{i < j < 
k} a _{i} a _{j} a _{k} \geq 0 $ (the equality holds when we have only 
two pairs since one can not have three different indices). Therefore we 
come to the conclusion that $ N _{2} N _{6}/ N _{4} ^{2} $ is 
minimized when we take a single pair of plane waves with wavevectors 
(${\bf  q} , -{\bf  q}$), corresponding to a simple order parameter 
proportional to $ \cos ({\bf  q}.{\bf  r})$, which has hence a planar 
symmetry. It is unfavourable to increase the number of plane waves. 
This is more easily seen in the particular situation where these $N$ 
plane waves have the same amplitude. In this case we merely have $ N 
_{2} N _{6}/ N _{4} ^{2} = [ 15 N ^{2} - 45 N + 40 ] / 9 (N-1) 
^{2}$ which increases regularly with increasing N, and is minimum for 
$N = 2$ . In this last case $ N _{2} N _{6}/ N _{4} ^{2} = 
\frac{10}{9}$. It is interesting to remark that this conclusion is 
opposite to what one would obtain by considering the fourth order term 
alone in Eq.(9) and omitting the sixth order one. Since the fourth order 
term grows (compared to the second order one) with the number $N$ 
plane waves, one would conclude that it is better to increase this 
number. The opposite turns out to be true because the sixth order term 
grows even faster with the number of plane waves. This shows quite 
clearly that, in contrast with what one might hope, the consideration of 
the fourth order term is not enough to conclude about the actual ground 
state of system.

\vspace{4mm} 
We find then explicitely $ A _{0} = \frac{5}{36} + 2.02 \:  10 ^{-3} 
$. We note that this is quite close to the standard FFLO transition itself. 
For comparison the standard Clogston-Chandrasekhar 
\cite{clog,chand} first order transition is given, as we have seen, by $ 
A _{0} = \frac{1}{8}$ so the difference with the standard FFLO 
transition is $ - 1.39  \:  10 ^{-2} $ which is also rather small. Since we 
know that the standard Clogston-Chandrasekhar transition and standard 
FFLO transition stay close beyond the vicinity of the TCP and that this 
proximity between these two lines extends down to zero temperature, it 
seems quite possible that the same is true for our first order transition 
line. This is indeed what has been found numerically by Matsuo et al. 
\cite{matsuo}. On the other hand we know from the work of Burkhardt 
and Rainer \cite{br} that the transition is second order in a 2D situation. 
So it is of some interest to consider formally an arbitrary dimension to 
see how one goes from the 2D to the 3D situation. For the case of our 
planar order parameter, there is no difficulty in working with an 
arbitrary dimension D. As we have mentionned at the end of section II, 
one has just to modify some coefficients in Eq.(7) : here one has to 
multiply the $ Q ^{2}$ terms by $3/D$ and the $ Q ^{4}$ terms by 
$15/D(D+2)$. Reproducing then the above analysis, we obtain that the 
FFLO transition occurs for $A _{0} =  \frac{D+2}{12 D}$ with a 
wavevector $\bar{q} ^{2} = \frac{D+2}{6}$. One finds again that, in 
order to obtain a phase transition higher than FFLO, the coefficient of 
the second term in the r.h.s. of Eq.(21) has to be negative. But this 
coefficient is now in general $ N _{4} (\frac{ \alpha (D+2)}{D}- 1)/4 $ 
with explicitely $ \alpha = \frac{5}{9} $. So we find $ D _{c} = 2.5 $ 
as the critical dimension to have a first order transition, which is 
halfway between the physical situations $D = 2$ and $D = 3$. Hence, 
in contrast to what one might think, the first order transition we have 
found is not 'very near to be second order', despite the proximity 
between the first order transition line and the standard second order 
FFLO line. The result Eq.(21) for the location of the transition for this 
planar state becomes :
\begin{eqnarray}
A _{0} = \frac{D+2}{12 D} + \frac{(2 D - 5 ) ^{2}}{15 D (7 D - 10)}  
\label{eq27}
\end{eqnarray}
with the value of $ \bar{\Delta}_{m}$ of $  \bar{\Delta}$ at the 
threshold given by :
\begin{eqnarray}
\bar{\Delta}^{2}_{m} = \frac{4}{5}\: \frac{2 D - 5}{7 D - 10} 
\label{eq28}
\end{eqnarray}
In two dimensions we can again make use of Eq.(20) to compare all the 
possible states. As we have seen the transition is second order and we 
can neglect the sixth order term as in the original LO analysis. The 
fourth order term is positive and, roughly speaking, below the 
transition the state with smaller fourth order term is selected. This leads 
to take $ N _{4} / N _{2} ^{2} = 3 (2-S _{2})/2 $ as small as 
possible. So we find again that the single pair state $ N = 2 $ is 
selected, which corresponds to the state investigated by Burkhardt and 
Rainer \cite{br}.

\vspace{4mm} 
In closing this section, let us note that we have have found a non-
degenerate minimum for the free energy, namely the planar state $ \cos 
({\bf  q}.{\bf  r})$ corresponding to two plane waves with opposite 
wavevectors and equal amplitude in Eq.(11). However by adding to 
this order parameter another cosine with a small amplitude, namely $ 
\epsilon  \cos ({\bf  q _{1}}.{\bf  r})$, we can have a free energy 
arbitrarily close to the planar state free energy. Conversely the location 
for the transition can be made arbitrarily close to the planar state 
transition. These states are highly degenerate since the relative 
directions of $ {\bf  q}$ and ${\bf  q _{1}}$ do not matter. These are 
some kind of excited states for our system with arbitrarily small energy 
and it is possible to speculate that they play a role in the physics of our 
system.

\section{BEYOND  THE  LO  SUBSPACE}

In the preceding section we have restricted to the LO subspace our 
search for the state with lowest free energy at the transition and 
therefore with highest critical temperature. While this restriction is 
justified when the free energy expansion is limited to the fourth order 
term (this term being treated as a perturbation), it is no longer valid 
when the sixth order term is included. And indeed the actual minimum 
state does not belong to the LO subspace, since it can be checked that 
Eq.(11) can not satisfy the Euler-Lagrange equation corresponding to 
Eq.(10) because of the non linear terms. However we have found in the 
preceding section a non degenerate minimum corresponding to the 
planar state. The selection of the wavevectors of this state was due to 
the fourth order term while the number of components was controled by 
the sixth order one, which is otherwise quite structureless. If the 
nonlinearities produce only a rather small change on the solution, it is 
reasonable to look for the actual solution as being 'near' the solution $ 
\cos ({\bf  q}.{\bf  r})$ we have found in section IV. In particular it is 
reasonable to look for an order parameter with a one-dimensional 
dependence on ${\bf  r}$ and we will also assume it to be real. And 
indeed we will find a solution which is quite near $ \cos ({\bf  q}.{\bf  
r})$. Although this argumentation is only a self-consistent one (we can 
not exclude that a strong nonlinear modification of a three-dimensional 
solution produces the actual minimum), we note that the numerical 
exploration of Houzet \textit{et al} \cite{buz1} has indeed produced a 
real one-dimensional order parameter for the minimum.

\vspace{4mm} 
With a one-dimensional real order parameter the reduced free energy 
simplifies into :
\begin{eqnarray}
F  = \int  dx \: [A _{0} \bar{\Delta}^{2} -  \frac{1}{3} 
\bar{\Delta}^{'2} + \frac{1}{5}  \bar{\Delta}^{''2}] 
-  \int dx \: [\frac{1}{4}  \bar{\Delta}^{4} - \frac{5}{6} 
\bar{\Delta}^{2}\bar{\Delta}^{'2}]  +   \frac{1}{8}  \int dx \: 
\bar{\Delta}^{6}
\label{eq29}
\end{eqnarray}
where $\bar{\Delta}'$ and  $\bar{\Delta}''$ are first and second 
derivative of $\bar{\Delta}$ with respect to $x$. Although we deal with 
a nonlinear problem we can still minimize with respect to the amplitude 
of the order parameter, just as we have done at the end of the preceding 
section (this works actually also for a three-dimensional order 
parameter). We set $ \bar{\Delta}(x) = a \delta (x)$, where $\delta (x)$ 
is a normalized spatial function (for example by $\int \delta ^{2} = 1$ ). 
$\delta (x)$ gives the shape of the order parameter while $a$ 
corresponds to its amplitude. When we substitute in Eq.(29) we obtain 
for $F/ a ^{2}$ a quadratic form in $a ^{2}$. Writing again that, for a 
specific function $ \delta (x)$, this form has double root at the transition 
leads us to an expression for $ A _{0}$ which does not depend 
anymore on the normalization of $\delta $ :
\begin{eqnarray}
A _{0} = \frac{[ \int \delta^{4} -  \frac{10}{3} (\delta \delta ')^{2}] 
^{2}}{8 \int \delta ^{2} \int \delta ^{6}}  + \frac{ \int \frac{1}{3}  
\delta^{'2} -  \frac{1}{5} \delta ^{''2}}{\int \delta ^{2}}
\label{eq30}
\end{eqnarray}
and we want now to maximize A$_0$. Similarly the condition that the 
free energy is zero at the transition imply :
\begin{eqnarray}
\int \bar{\Delta}^{6} = \int \bar{\Delta}^{4} -  \frac{10}{3} 
\bar{\Delta} ^{2} \bar{\Delta}^{'2}
\label{eq31}
\end{eqnarray}
or equivalently :
\begin{eqnarray}
A _{0} \int \bar{\Delta}^{2} = \int  \frac{1}{8}\bar{\Delta}^{6}+ 
\frac{1}{3} \bar{\Delta}^{'2} -  \frac{1}{5}  \bar{\Delta}^{''2}
\label{eq32}
\end{eqnarray}
which gives the amplitude of the order parameter.

\vspace{4mm} 
The condition that $ A _{0}$ is maximum can be obtained from 
Eq.(30) but it is easier to deduce it from Eq.(29). One finds the 
ordinary nonlinear differential equation :
\begin{eqnarray}
\frac{1}{5} \bar{\Delta}'''' + (\frac{1}{3}- 
\frac{5}{6}\bar{\Delta}^{2}) \bar{\Delta}'' -  \frac{5}{6}\bar{\Delta} 
\bar{\Delta}^{'2} + \frac{3}{8} \bar{\Delta}^{5} -  \frac{1}{2} 
\bar{\Delta}^{3} + A _{0} \bar{\Delta} = 0
\label{eq33}
\end{eqnarray}
(analytically this equation can be integrated once and then reduced to a 
second order nonlinear differential equation but this is of no real help 
for numerics). One checks readily that the cosine form of the order 
parameter $ \bar{\Delta}(x) = a \cos (\bar{q}x)$ is not a solution. It 
satisfies only the linear part of the equation, which leads to A$_0 = 
\bar{q}^2 /3 - \bar{q}^4 /5$. Looking for the maximum of $ A _{0}$ 
gives the FFLO result $ A _{0}= 5/36 \simeq 0.1389$ with $ \bar{q} 
_{FFLO}^2 = 5/6$. This FFLO solution corresponds to keep only the 
second term in the r.h.s. of Eq.(30). This term produces for $ A _{0}$ 
a kind of (inverted) effective potential $ \bar{q}^2 /3 - \bar{q}^4 /5$ 
which has a strong maximum for the FFLO solution. When we write a 
Fourier expansion of the solution this potential selects quite effectively 
the wavevectors in the close vicinity of the FFLO result $ \bar{q} 
_{FFLO}$.

\vspace{4mm} 
Indeed the numerical exploration of Eq.(33) for $ A _{0}$ very close to 
the maximum gives for the solutions one or two basis frequencies 
$q_0$ and $q_1$ which are close to $ \bar{q} _{FFLO}$. The other 
frequencies appearing in a Fourier analysis are simply odd 
combinations of  $q_0$ and $q_1$ like $2 q_0 \pm q_1$, $2 q_1 \pm 
q_0$, $3 q_0$, $4 q_1 \pm q_0$, etc. The weights of these frequencies 
depend on their order, i.e. the higher the frequency, the smaller the 
weight. This shows explicitely the strength of the effective FFLO 
potential in A$_0$ (the second term in the r.h.s. of Eq.(30)). We have 
investigated analytically the efficiency of a small frequency splitting by 
writing $q_1 = q_0 + \epsilon$ with $\epsilon \to 0$ (but $\epsilon \ne 
0$ because this limit is singular). However a single frequency $q_1 = 
q_0 $ turned out to be always better. We are finally lead to the 
conclusion that going beyond the LO subspace produces only small 
corrections due to nonlinearities.
These corrections correspond to odd harmonics of the fundamental 
frequency $\bar{q}$ and they are small because of the efficiency of the 
effective FFLO potential. Actually if we assume from the start that the 
corrections to $\delta _{0}(x) = \cos (\bar{q}x)$ are small, we can 
write $\delta (x) = \delta _{0}(x) + \delta _{1}(x)$, with $\delta 
_{1}(x)$ small, and perform a first order expansion of Eq.(30). One 
finds readily that it is most favorable to take $\delta _{1}(x)$ 
proportional to $ \cos ^{3}(\bar{q}x)$ (another contribution from $ 
\cos ^{5}(\bar{q}x) $ is quite small), but one has to go to second order 
to find the amplitude. Actually, once it is proved that this harmonic 
expansion is the correct answer, it is much easier to minimize Eq.(30) 
numerically which avoids cumbersome calculations. Specifically we 
considered the trial function $\delta (x)  = \cos(\bar{q}x) + a_3 \cos(3 
\bar{q} x + \phi_1) + a_5 \cos ( 5  \bar{q} +\phi_2)$ and numerically 
maximized A$_0$ (some more complete trial functions that we also 
tried eventually reduced to this form when maximized). We found $a_3 
= -1.33.10^{-2}$, $a_5 = 1.62.10^{-4}$, $\bar{q} = 0.793$, 
$\phi_{1,2} = 0$ and $A_0 = 0.141604$.  This form is very close to 
the cosine solution and the improvement in A$_0$ compared to our 
result Eq.(22) for the cosine solution $A_0 = 31/220 \simeq 0.14091$ 
is pretty small in absolute values, although it gives a significant increase 
of $7 \:  10 ^{-4} $ to our gain of $2.02 \:  10 ^{-3} $ compared to the 
FFLO result. Our result $\bar{q} = 0.793$ has to be compared to our 
result in LO subspace from Eq.(23) $\bar{q} = 0.829$ and to $\bar{q} 
_{FFLO} = 0.913 $.

\section{ CONCLUSION }

In this paper we have explored analytically the nature of the transition to 
the FFLO superfluid phases in the vicinity of the tricritical point, where 
these phases begin to appear. This region is convenient for the 
analytical study we make because, in the vicinity of this point, one can 
make use of an expansion of the free energy up to sixth order, both in 
order parameter amplitude and in wavevector. Despite this 
simplification one has still a complex nonlinear problem to solve. We 
have first explored the minimization of this free energy within the LO 
subspace, made of arbitrary superpositions of plane waves. We have 
seen that the standard second order FFLO phase transition is unstable 
and that a first order transition occurs at higher temperature. Within this 
subspace we have shown that it is favorable to have a real order 
parameter. Then among these states we have shown that those with the 
smallest number of plane waves are prefered. This leads to retain only 
two plane waves, corresponding to an order parameter with a $\cos({\bf  
q}_{0}. {\bf  r})$ dependence, in agreement with preceding work 
\cite{buz1}. Finally we have shown that, when releasing the constraint 
of working within the LO subspace, the order parameter at the transition 
is only very slightly modified by higher harmonics contributions and 
we have been able to ascribe this result to the strong selection of the 
wavevector caused by the second order terms of the free energy, 
corresponding physically to the standard FFLO transition.

We are very grateful to X. Leyronas for stimulating discussions and to 
K. Nagai for making us aware of his work, during a meeting very 
nicely organized by G. Eska and H. Brandt for D. Rainer's 60th 
birthday.

\vspace{4mm} 
* Laboratoire associ\'e au Centre National
de la Recherche Scientifique et aux Universit\'es Paris 6 et Paris 7.


\begin{references}
\bibitem{ff}P. Fulde and R. A. Ferrell, Phys.Rev. {\bf 135}, A550 
(1964).
\bibitem{larkov}A. I. Larkin and Y. N. Ovchinnikov, ZhETF {\bf 47}, 
1136 (1964) [Sov. Phys. JETP {\bf 20}, 762 (1965)].
\bibitem{singleton}J. Singleton, J. A. Symington, M. S. Nam, A. 
Ardavan, M. Kurmoo and P. Day, cond-mat/0105335.
\bibitem{gloos}K. Gloos et al., Phys.Rev.Lett.{\bf 70},501 (1993).
\bibitem{marcojin}B. DeMarco and D. S. Jin, Science {\bf  285}, 
1703 (1999).
\bibitem{trusc}A. G. Truscott et al., Science {\bf  291} 2570 (2001).
\bibitem{salomon} F. Schreck et al., Phys.Rev.Lett.{\bf 87} 080403 
(2001).
\bibitem{stoofal}  H. T. C. Stoof, M. Houbiers, C. A. Sackett and R. 
G.Hulet, Phys. Rev. Lett. {\bf 76}, 10 (1996).
\bibitem{rc}R. Combescot, Phys.Rev.Lett.{\bf 83}, 3766 (1999).
\bibitem{bkk}M. A. Baranov, M. Yu. Kagan and Yu. Kagan, JETP 
Lett. {\bf  64}, 301 (1996).
\bibitem{rcffan}R. Combescot, Europhys. Lett.{\bf 55}, 150 (2001).
\bibitem{buz1}M. Houzet, Y. Meurdesoif, O. Coste and A. I. Buzdin, 
Physica C {\bf  316} 89 (1999).
\bibitem{matsuo}S. Matsuo, S. Higashitani, Y. Nagato and K. Nagai, 
J. Phys. Soc. Japan, {\bf  67} 280 (1998).
\bibitem{br}H. Burkhardt and D. Rainer, Ann.Physik {\bf  3}, 181 
(1994).
\bibitem{buz2}A. I. Buzdin and H. Kachkachi, Phys. Lett. A {\bf  
225} 341 (1997).
\bibitem{agd}A. A. Abrikosov, L. P. Gorkov and I. E. Dzyaloshinski, 
Methods of Quantum Field Theory in Statistical Physics, Prentice-Hall 
(1963).
\bibitem{eil}G. Eilenberger, Z. Phys. {\bf  182} 427 (1965).
\bibitem{gork}L. P. Gorkov, ZhETF {\bf 36}, 1918 (1959) [Sov. 
Phys. JETP {\bf 9}, 1364 (1959)].
\bibitem{note}Actually this is only correct for the generic situation 
where the $\{ {\bf  q}_{i} \}$ have no other specific relation than being 
associated in pairs $({\bf  q}_{i}, -{\bf  q}_{i})$. A simple case where 
additional degeneracy occurs is the set of 8 wavectors $ (\pm 1, \pm  1, 
\pm  1) / \sqrt{3} $, which has a cubic symmetry. In this case we find 
for example $ N _{4} = 216$ instead of $168$. The same caveat is 
valid for Eq.(24) and Eq.(25) below. However we will see quite 
generally that it is unfavorable to increase the number of plane waves in 
order to minimize the free energy. Since these exceptional cases imply a 
large number of plane waves, they should have higher free energy and 
we do not expect them to alter our final result.
\bibitem{clog}A. M. Clogston, Phys.Rev.Lett.{\bf 9}, 266 (1962).
\bibitem{chand}B. S. Chandrasekhar, Appl.Phys.Lett. {\bf 1}, 7 
(1962).
\end{references}
\end{document}